\documentclass[onecolumn,prd,aps,amsmath,showpacs,amssymb]{revtex4}
\advance\oddsidemargin by -\textwidth
\advance\oddsidemargin by -1in
\evensidemargin=\oddsidemargin
\def\fsl#1{\setbox0=\hbox{$#1$}           
   \dimen0=\wd0                                 
   \setbox1=\hbox{/} \dimen1=\wd1               
   \ifdim\dimen0>\dimen1                        
      \rlap{\hbox to \dimen0{\hfil/\hfil}}      
      #1                                        
   \else                                        
      \rlap{\hbox to \dimen1{\hfil$#1$\hfil}}   
      /                                         
   \fi}    
\makeatletter
\def\@maketitle{\newpage
 \null
 {\normalsize \tt \begin{flushright} 
  \begin{tabular}[t]{l} \@date 
  \end{tabular}
 \end{flushright}}
 \begin{center}
 \vskip 2em
 {\LARGE \@title \par} \vskip 2.5em {\large \lineskip .5em
 \begin{tabular}[t]{c}\@author 
 \end{tabular}\par} 
 \end{center}
 \par
 \vskip 1.5em} 
\makeatother
\usepackage{graphicx}
\usepackage{subfigure}
\usepackage{color}
\usepackage{dcolumn}
\usepackage{bm}
\usepackage{latexsym}
\usepackage{amsmath}
\usepackage{amsfonts}
\usepackage{amssymb}

\newcommand{\bra}[1]{\langle #1 |}
\newcommand{\ket}[1]{| #1 \rangle}

\newcommand{\TeV}{\, \text{TeV}}
\newcommand{\GeV}{\, \text{GeV}}

\newcommand{\gtwo}{I\kern-.1em I\,}
\newcommand{\be}{\begin{equation}}
\newcommand{\ee}{\end{equation}}
\newcommand{\bea}{\begin{eqnarray}}
\newcommand{\eea}{\end{eqnarray}}
\newcommand{\bpm}{\begin{pmatrix}}
\newcommand{\epm}{\end{pmatrix}}
\newcommand{\cl}{\, \rm C.L.}
\newcommand{\1}{I}
\newcommand{\2}{I\hspace{-.1em}I}
\newcommand{\3}{I\hspace{-.1em}I\hspace{-.1em}I}

\begin{document}
\title{Flavor constraints in a Bosonic Technicolor model}

\author{Hidenori S. Fukano$^{a,b}$}
\email{hidenori.f.sakuma@jyu.fi}
\author{Matti Heikinheimo$^c$}
\email{hmatti@yorku.ca} 
\author{Kimmo Tuominen$^{a,b}$}
\email{kimmo.i.tuominen@jyu.fi}
\affiliation{$^a$Department of Physics, University of Jyv\"askyl\"a, P.O.Box 35, FIN-40014 Jyv\"askyl\"a, Finland \\
$^b$Helsinki Institute of Physics, P.O.Box 64, FIN-00014 University of Helsinki, Finland\\
$^c$Department of Physics and Astronomy, York University, 4700 Keele Street, Toronto, ON, M3J 1P3 Canada}

\begin{abstract}
Flavor constraints in a bosonic Technicolor model are considered. We illustrate different sources for their origin, and emphasize in particular the role played by the vector states present in the Technicolor model. This feature is the essential difference in comparison to an analogous model with two fundamental Higgs scalar doublets.
\end{abstract}

\maketitle

\section{Introduction}

Technicolor \cite{Weinberg:1979bn,Susskind:1978ms} remains as one of the compelling alternatives to electroweak symmetry breaking extending the simple Higgs sector of the Standard Model (SM). Earliest models based on scaled up QCD run into trouble with precision data \cite{Peskin:1990zt}. Recent developments suggest that viable models are obtained via introduction of technifermions transforming under higher representations of the gauge group \cite{Sannino:2004qp,Dietrich:2005jn}. In this paper we study one of these new models, the so called Next to Minimal Walking Technicolor (NMWT).  In NMWT model the Technicolor gauge group is SU(3), and the technicolor matter is constituted by two Dirac flavors transforming under the two-index symmetric, i.e. the sextet representation of the gauge group. Phenomenology of this Technicolor theory has been studied in \cite{Belyaev:2008yj}. Its properties on the lattice have also been investigated \cite{Shamir:2008pb,DeGrand:2008kx,DeGrand:2010na,Fodor:2009ar,Kogut:2010cz}.

The Technicolor interaction only explains the mass patterns in the gauge sector of the Standard Model (SM), and to explain the fermion mass patterns and hierarchies extensions are needed. One possibility is so-called Extended Technicolor (ETC) \cite{Dimopoulos:1979es,Eichten:1979ah}, where one imagines the technifermions and ordinary fermions to be combined under a single representation of a larger gauge group whose breaking then induces effective low-energy four-fermion interaction terms. These, in turn, lead to mass terms for SM fermions as the technifermions condense. However, alternative avenues for addressing the fermion mass generation in Technicolor framework do exist. Instead of assuming the existence of an ETC sector, one may reintroduce scalars with renormalizable Yukawa interactions with ordinary and technicolored matter fields. This is so called bosonic Technicolor \cite{Simmons:1988fu,Samuel:1990dq,Kagan:1991gh,Carone:1992rh,Hemmige:2001vq}. To naturalize the scalars, one can supersymmetrize Technicolor \cite{Dine:1981za,Dobrescu:1995gz,Antola:2010nt}.

Generally, when considering any of the above mentioned extensions of a technicolor model one must pay attention to possibly large contributions to the  flavor changing neutral current (FCNC) processes. On the other hand, the fact that already the underlying Technicolor theory contributes to flavor physics has received less attention in the literature \cite{Fukano:2009zm}. This is so since in Technicolor, the spin-one composite objects and the electroweak gauge bosons with identical quantum numbers will mix providing additional contributions to the flavor observables via the usual box diagrams. So, given a Technicolor model and its bosonic extension, one will be able to constrain both sectors via flavor physics.

In this paper we consider a bosonic technicolor model obtained by coupling the NMWT model with a SM-like scalar doublet \cite{Antola:2009wq}. We do not specify if this is fundamental scalar, possibly a low energy remnant of some supersymmetric scenario, or a composite with the compositeness scale much higher than the energies where the phenomenology of this theory is studied. With a bottom-up model building attitude we simply treat the effective Lagrangian, to be introduced in the next section, as a low energy effective theory and study its associated phenomenology.

We will derive the results required for the analysis of the flavor constraints in this model, but the developed formalism can be applied to other model setups extending beyond basic Technicolor models. In addition, another source of constraints we consider is provided by the precision electroweak data, i.e. the oblique corrections. As we will see, they constrain the spectrum of the bosonic Technicolor in an interesting way, and also mix with the flavor constraints due to the Weinberg sum rules.

As a conclusion, we find that the bosonic NMWT model is viable in light of the present data. However, the constraints do impose severe restrictions on the parameter space of the model. We will determine the results on the masses and mixing patterns of the states. In particular, the compatibility with the precision data inescapably leads to the prediction that at least one of the neutral scalars in the model has to be light, with a mass below 200 GeV. This gives an important phenomenological handle on the model with respect to the LHC data even if the couplings of the lightest scalar are weaker than those of the SM Higgs. 
 
\section{Bosonic Next to Minimal Walking Technicolor}

The low energy effective Lagrangian for Bosonic NMWT has been introduced in detail in \cite{Antola:2009wq}, but for completeness we recall the basic results here. The effective Lagrangian is, schematically,
\bea
{\cal L} =
{\cal L}^{\rm eff}_{\rm TC}|_{{\rm Higgs}=0} + {\cal L}_{\rm SM}|_{{\rm Higgs}=0}
+ {\cal L}_{\rm Higgs} + {\cal L}_{\rm Yukawa}\,.
\label{BTCLagrange}
\eea
Here ${\cal L}^{\rm eff}_{\rm TC}|_{{\rm Higgs}=0}$ is almost the same as the effective theory of the NMWT constructed in \cite{Belyaev:2008yj,Appelquist:1999dq,Foadi:2007ue}. The only difference is that ${\cal L}^{\rm eff}_{\rm TC}|_{{\rm Higgs}=0}$ above does not include the {\it composite} higgs scalar explicitly, i.e. ${\cal L}^{\rm eff}_{\rm TC}|_{{\rm Higgs}=0}$ has only the electroweak gauge bosons and the vector mesons in the sense of the generalized hidden local symmetry \cite{Bando:1987ym,Bando:1987br}. 
${\cal L}_{\rm SM}|_{{\rm Higgs}=0}$ contains the contribution of the SM degrees of freedom excluding the fundamental Higgs scalar doublet. Finally, 
${\cal L}_{\rm Higgs}$ is 
\bea
{\cal L}_{\rm Higgs} = |D_\mu M|^2 + |D_\mu H|^2 + ( M ,H\text{-mixing term}) -V(M,H)\,,
\label{L-higgs}
\eea
where $M$ is the {\it composite} scalar field formed by the walking
technicolor dynamics and  $H$ is another scalar field. As discussed in
the introduction we do not specify if $H$ is a fundamental or a composite object.
In general $M,H$ are technicolor singlets and have the same EW-charge under $SU(2)_L \times U(1)_Y$ as the SM-Higgs field. Hence, due to the underlying symmetry the $H$ and $M$ fields mix with each other and this is accounted for by the $(M,H\text{-mixing term})$ in Eq.(\ref{L-higgs}). This mixing is resolved via transformation \cite{Antola:2009wq} 
\bea
\bpm M \\ H \epm
=
\bpm A & B \\ -A & B \epm
\bpm M_- \\ M_+ \epm\,,
\label{mhtransf}
\eea
where 
\bea
A=\frac{1}{\sqrt{2}}\frac{1}{\sqrt{1-\frac{c_3}{\alpha}y_Q}},\quad
B=\frac{1}{\sqrt{2}}\frac{1}{\sqrt{1+\frac{c_3}{\alpha}y_Q}}.
\label{AB}
\eea
The parameter $c_3$ is ${\mathcal O}(1)$ low energy constant, $\alpha=\Lambda/ f\sim 4\pi$ is the ratio between the mass scale of the lightest non-Goldstone mode and the Goldstone boson decay constant $f$.  Finally, $y_Q$ is the Yukawa coupling between techniquarks and the scalar doublet $H$. These parameters originate from the underlying Lagrangian, for details, see \cite{Antola:2009wq}. Note that since the mixing of $M$ and $H$ occurs also on the level of the kinetic term, the transformation in (\ref{mhtransf}) is not simply a rotation.

After the kinetic terms in Eq.(\ref{L-higgs}) are diagonalized
\bea
{\cal L}_{\rm Higgs} = |D_\mu M_+|^2 + |D_\mu M_-|^2 -V(M_\pm)\,,
\eea
the physical mass eigenstates of the scalar sector are obtained via an additional rotation. We rewrite $M_\pm$ as 
\bea
\bpm G \\ \Pi \epm
=
\bpm \cos \beta & \sin \beta \\ -\sin \beta & \cos \beta \epm
\bpm M_- \\ M_+ \epm\,,
\eea
where $G$ indicates the would-be Nambu-Goldstone boson part which is absorbed in the longitudinal mode of the EW-gauge bosons and $\Pi$ indicates the physical part which will remain in the spectrum.
Thus after combining these two transformations, $M,H$ can be represented by $G,\Pi$ as 
\bea
\bpm M \\ H \epm
=
\bpm Ac_\beta+Bs_\beta & -As_\beta+Bc_\beta \\ -Ac_\beta+Bs_\beta & As_\beta+Bc_\beta \epm
\bpm G \\ \Pi \epm\,,
\eea
where we defined $c_\beta\equiv\cos\beta$ and $s_\beta\equiv\sin\beta$.

We parametrize the EW-doublets $G,\Pi$ as
\bea
G = 
\bpm 
i G^+ \\[2ex] \dfrac{v_{\rm EW} + h^0 - iG^0}{\sqrt{2}}
\epm
\quad,\quad
\Pi = 
\bpm 
i \pi^+ \\[2ex] \dfrac{ H^0 - i\pi^0}{\sqrt{2}}
\epm\,,
\eea
where $M^2_W = g^2_{\rm EW} v^2_{\rm EW}/4$ in which $g_{\rm EW}$ is the physical $SU(2)_L$ gauge coupling. 

The final contribution in (\ref{BTCLagrange}), ${\cal L}_{\rm Yukawa}$,
includes the SM Yukawa sector. The Yukawa couplings of the SM quarks are
given by
\bea
{\cal L}_{{\rm Yukawa},q} = 
-\tilde{Y}^{ij}_d \bar{\tilde{q}}^i_L H \tilde{d}^j_R - \tilde{Y}^{ij}_u \bar{\tilde{q}}^i_L (i \sigma_2 H^*) \tilde{u}^j_R + \rm{h.c.}\,,
\eea
where we assume that only $H$ has Yukawa coupling with the SM fermions (denoted with a tilde) in the weak gauge eigenbasis. The parameters $\tilde{Y}_{u,d}$ are the Yukawa matrices and have non-diagonal components in general. 
In terms of the physical mass eigenstates the scalar field $H$ is represented in the unitary gauge as
\bea
H = 
(-Ac_\beta+Bs_\beta) 
\bpm 
0 \\[2ex] \dfrac{ v_{\rm EW} + h^0}{\sqrt{2}}
\epm
+
(As_\beta+Bc_\beta)
\bpm 
i \pi^+ \\[2ex] \dfrac{ H^0- i \pi^0}{\sqrt{2}}
\epm\,.
\label{higgs-u-gauge}
\eea

Now we divide ${\cal L}_{{\rm Yukawa},q}$ into three parts as
\bea
{\cal L}_{{\rm Yukawa},q} = {\cal L}_{\rm Yukawa} ({\rm mass})+ {\cal L}_{\rm Yukawa}(h^0,H^0) + {\cal L}_{\rm Yukawa}(\pi^\pm)\,,
\label{divide-yukawa}
\eea
where,  
the first term becomes fermion mass term after changing from the weak basis to the mass eigenbasis for fermions, the second one includes physical neutral higgs and the third one includes physical charged pions. Since only one of the Higgs doublets couples to fermions, the tree level contributions to FCNC interactions are absent in this model, and in our analysis of flavor constraints it suffices to concentrate on the first and third terms of (\ref{divide-yukawa}). The first term in Eq.(\ref{divide-yukawa}) is represented as
\bea
{\cal L}_{\rm Yukawa} ({\rm mass}) 
\!\!&=&\!\!
- \frac{v_{\rm EW} (-Ac_\beta+Bs_\beta)}{\sqrt{2}} 
\left[ 
\bar{\tilde{d}}^i_L \tilde{Y}^{ij}_d  \tilde{d}^j_R - \bar{\tilde{u}}^i_L \tilde{Y}^{ij}_u \tilde{u}^j_R + \rm{h.c.} 
\right] \nonumber\\
\!\!&=&\!\!
- \frac{v_{\rm EW} (-Ac_\beta+Bs_\beta)}{\sqrt{2}} 
\left[ 
\bar{d}^i_L Y^{ii}_d  d^i_R - \bar{u}^i_L Y^{ii}_u u^i_R + \rm{h.c.}
\right]\,,
\eea
where on the second line the SM fermions without tilde are in the fermion mass eigenbasis and the fermion mass matrices are given by
\bea
m^{ii}_{u(d)} = \frac{v_{\rm EW}  (-Ac_\beta+Bs_\beta)}{\sqrt{2}} 
\bpm Y^{11}_{u(d)} & \, & \, \\ \, & Y^{22}_{u(d)} & \, \\ \, & \, & Y^{33}_{u(d)} \epm
\equiv
\bpm m_{u(d)} & \, & \, \\[1.5pt] \, & m_{c(s)} & \, \\[1.5pt] \, & \, & m_{t(b)} \epm\,.
\eea
In the fermion mass eigenbasis, the third term in Eq.(\ref{divide-yukawa}) is 
\bea
{\cal L}_{\rm Yukawa}(\pi^\pm) 
\!\!&=&\!\!
-i \pi^+ (As_\beta+Bc_\beta) 
\left[ \bar{u}^i_L Y^{jj}_d d^j_R - \bar{u}^i_R Y^{ii}_u d^j_L \right] V_{ij}
+ \rm{h.c.} \nonumber\\
\!\!&=&\!\!
-i \pi^+ X 
\left[ \bar{u}^i_L m^{jj}_d d^j_R - \bar{u}^i_R m^{ii}_u d^j_L \right] V_{ij}
+ \rm{h.c.},
\label{L-ffpi}
\eea
where we defined
\bea
X\equiv\frac{As_\beta+Bc_\beta}{-Ac_\beta+Bs_\beta}.
\label{Xvariable}
\eea 

As is evident from (\ref{L-ffpi}), the constraints will be most conveniently imposed on the parameter $X$. This will also hide the parameters of the underlying theory defined briefly below Eq. (\ref{AB}) from the analysis into a single quantity.

\section{Oblique corrections}
\label{obliques}

The oblique corrections $S$ and $T$ \cite{Peskin:1990zt} are defined as
\bea
S &=& -16\pi \Pi_{3Y}^\prime(0),\nonumber \\
T &=& \frac{4\pi}{s_w^2c_w^2 M_Z^2}(\Pi_{11}(0)-\Pi_{33}(0)). 
\eea
To evaluate the corrections we need to estimate the new contributions to the vacuum polarizations
\be
\Pi_{3Y}(q^2),~~\Pi_{ii}(q^2) ~~(i=1,2),
\ee
arising from pion and scalar loop diagrams. The relevant diagrams and
Feynman rules following from the expansion of (\ref{L-higgs}) are given in \cite{Antola:2009wq}, and we will not repeat
the calculation in full here. The reader should note, however, that the
definition for the mixing angle $\theta$ used in \cite{Antola:2009wq}
differs from our $\beta$. The angles are related by
\be
\beta=\frac{\pi}{4}-\theta.
\ee
Taking that into account, we can use the results provided in
\cite{Antola:2009wq}. Generally, the $S$ parameter is given by
\bea
S = S_{V,A} + S_S\,,
\eea
where $S_{V,A}$ indicates the contribution from the vector mesons $V,A$ and 
$S_S$ indicates the contribution from the scalar mesons $\pi$ and the Higgs bosons. Here we will only consider the contribution of the
scalar and pseudoscalar degrees of freedom, i.e. $S_S$. The contribution of the
vector mesons is discussed in section \ref{numresults}.

The origin of the $(S,T)$-plane corresponds to the SM with a given value
of the mass of the Higgs boson denoted by $m_{\textrm{ref}}$. We have removed the SM Higgs sector and added new sectors as described in the previous section. Thus the S-parameter is
\be
S=S_{\textrm{SM}}(m_{\textrm{ref}})-S_H(m_{\textrm{ref}})+S_{\textrm{new}}=S_{\textrm{new}}-S_H(m_{\textrm{ref}}),
\ee
because $S_{\textrm{SM}}(m_{\textrm{ref}})=0$ by definition. Similar
considerations apply to $T$, and using these definitions we obtain
finite expressions for $S$ and $T$. We use dimensional regularization in
the $\overline{MS}$ scheme, and find that the final result can be expressed in terms of two integrals,
\bea
I_1(m_1,m_2,q) &=& \int_0^1 dx\Delta\log\frac{\mu^2}{\Delta},\nonumber \\
I_2(m_1,m_2,q) &=& \int_0^1 dx m_1^2\log\frac{\mu^2}{\Delta},
\eea
where $\Delta\equiv\Delta(m_1,m_2,q)=xm_2^2+(1-x)m_1^2-x(1-x)q^2$ and $\mu$ is an arbitrary mass scale. Furthermore, for notational convenience we define
\bea
C_h &=& \frac{1}{v_{\rm EW}}(f_+\cos\beta-f_-\sin\beta),\nonumber \\
C_s &=& \frac{1}{v_{\rm EW}}(f_-\cos\beta+f_+\sin\beta),
\label{oblique_constants}
\eea
where $f_+=(f+v)/(2B)$, $f_-=(f-v)/(2A)$. The parameters $A$ and $B$
were defined in (\ref{AB}) and $f$ and $v$ are the vacuum expectation
values of the fields $M$ and $H$, respectively. Similar quantities, $f_+$ and $f_-$, referring to the mass eigenbasis of the scalars are related to the electroweak scale $v_{\rm EW}$ by $f_+^2+f_-^2=v_{\rm EW}^2$, so that $C_h^2+C_s^2=1$.


With these preliminary definitions we have
\bea
\Pi_{3Y}(q^2) &=& \frac{1}{32\pi^2}\left[-C_h^2\left(I_1(m_\pi,m_{h^0})-2I_2(M_Z,m_{H^0})+I_1(M_z,m_{H^0})\right)\right.\nonumber \\
& & -C_s^2\left(I_1(m_\pi,m_{H^0})-2I_2(M_Z,m_{h^0})+I_1(M_Z,m_{h^0})\right)\\
& & \left.+I_1(m_\pi,m_\pi)+I_1(M_Z,m_{\textrm{ref}})-2I_2(M_Z,m_{\textrm{ref}})\right],\nonumber
\eea
where we have dropped all $q$-independent contributions since we need only the $q^2$-derivative of this quantity. Note that here and later $m_{h^0}$ and $m_{H^0}$ are the masses of the physical scalars $h^0$ and $H^0$. The contributions to the correlators needed for the $T$-parameter are obtained similarly, and the result is
\bea
\Pi_{11}(q^2) &=& \frac{1}{32\pi^2}\left[C_h^2\left(I_1(M_W,m_{H^0})+I_1(m_\pi,m_{h^0})-2I_2(M_W,m_{H^0})\right)\right.\nonumber \\
& & +C_s^2\left(I_1(M_W,m_{h^0})+I_1(m_\pi,m_{H^0})-2I_2(M_W,m_{h^0})\right)\nonumber \\
& & +I_1(m_\pi,m_\pi)-I_1(M_W,m_{\textrm{ref}})+2I_2(M_W,m_{\textrm{ref}})\\
& & \left.+\frac{3}{2}m_\pi^2-\frac{1}{2}m_{\textrm{ref}}^2+\frac{1}{2}m_{h^0}^2+\frac{1}{2}m_{H^0}^2-\frac{1}{3}q^2\right],\nonumber
\eea
and
\be
\Pi_{33}(q^2)=\Pi_{11}(q^2) \quad \text{with}\quad M_W\rightarrow M_Z.
\ee

The results of a numerical computation and the corresponding constraints due to the data on $S$ and $T$ will be presented in Sec. \ref{numresults}.

\section{Flavor constraints}
\label{flavor}

In \cite{Fukano:2009zm}, the contributions to the flavor observables from the vector mesons for the NMWT model have been estimated.  
The analysis of \cite{Fukano:2009zm} was carried out in terms of the weak eigenstates of the vector mesons. In the present paper, we reconsider the analysis of \cite{Fukano:2009zm} in the mass basis and extend it to the case of bosonic NMWT. 
For this purpose, we should first determine the eigenvalues and eigenstates for the mass matrix of the vector bosons. In this paper, we take into consideration $\Delta F = 2$ constraints, so we concentrate on the charged vector bosons sector. Here $F$ labels the flavor relevant for each of the processes we consider. 
Thus we can represent the charged current coupling to the fermions as
\bea
&&
\frac{g}{\sqrt{2}} 
\left[ 
V^*_{ij} \cdot \tilde{W}^-_\mu  \bar{d}^j_L \gamma^\mu u^i_L + {\rm h.c.}
\right] \nonumber\\[1ex]
&&= 
\frac{g}{\sqrt{2}} 
\left[ 
V^*_{ij} \cdot \left( x_W W^-_\mu + x_V V^-_\mu + x_A A^-_\mu\right) 
\bar{d}^j_L \gamma^\mu u^i_L + {\rm h.c.}
\right]\,,
\label{L-ffg}
\eea
where the tilde notation refers to the weak basis and the fields and parameters without tilde refer to the mass basis.  The CKM matrix is denoted by $V$. The notations and complete expressions for the mass eigenvalues and eigenvectors of vector mesons are given in Appendix \ref{masseigenvalues}.

\begin{figure}
\begin{center}
\begin{tabular}[b]{lc}
\raisebox{-35pt}{(\1)} & \includegraphics[scale=0.65,angle=-90]{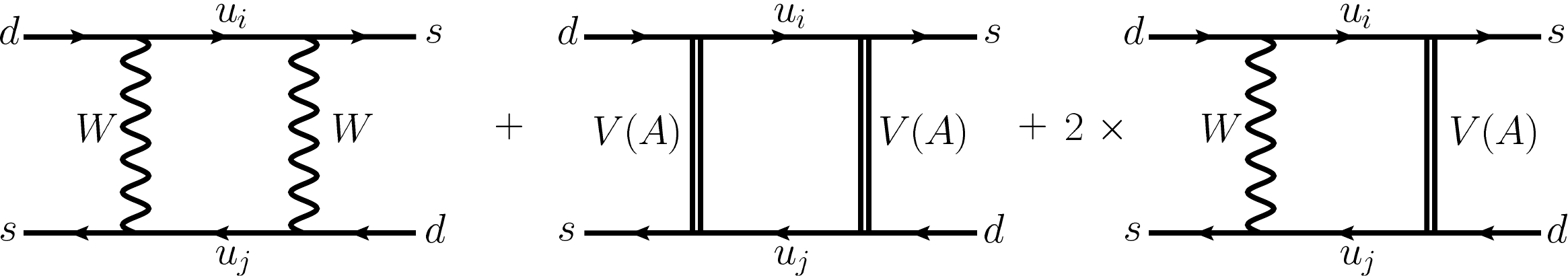} \\
\raisebox{-35pt}{(\2)} & \includegraphics[scale=0.65,angle=-90]{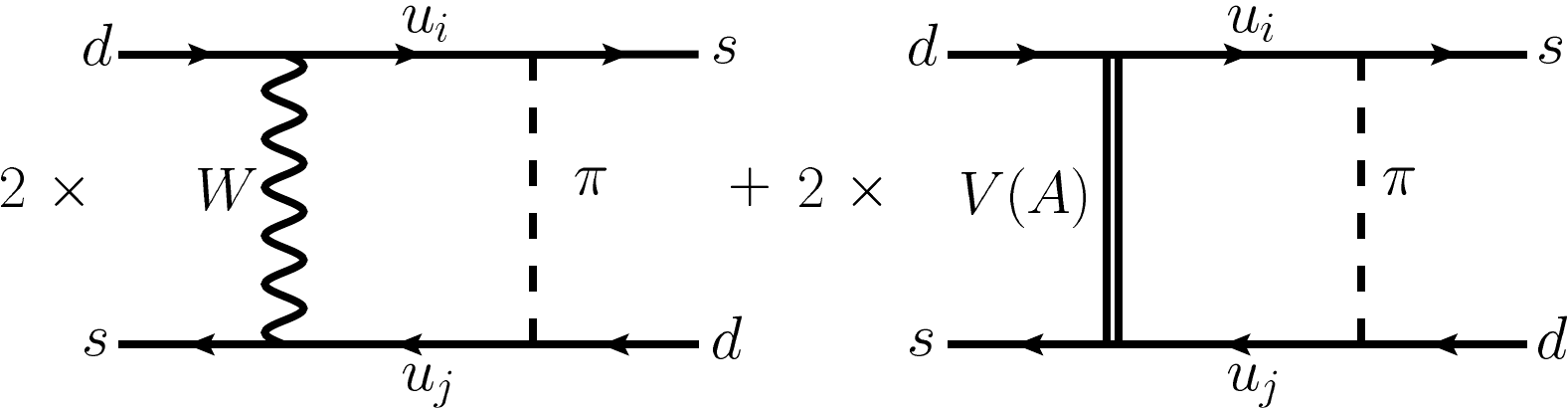} \\
\raisebox{-35pt}{(\3)} & \includegraphics[scale=0.65,angle=-90]{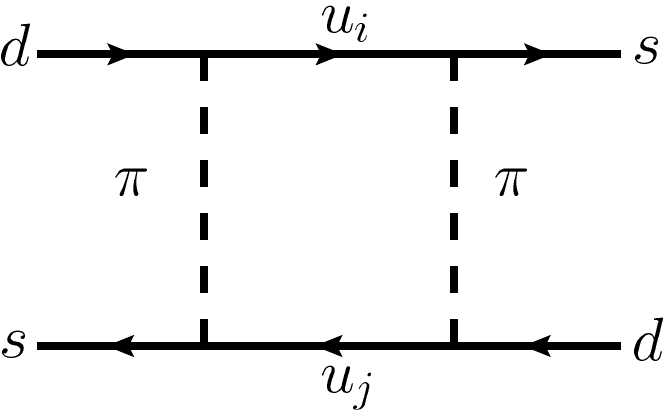} \\
\end{tabular}
\end{center}
\caption{Box diagrams for $\Delta S = 2$ scattering process in the unitary gauge.
To obtain $\Delta B = 2$ process, we should simply rename $s$ with $b$ and $d$ with $q\,(q=d,s)$ in the various diagrams. 
\label{FCNC-box}}
\end{figure}

To analyse the flavor changing neutral current interactions in the bosonic NMWT model, we compute the Box diagrams contributing to the $\Delta F = 2$ processes. The required diagrams are shown in Fig. \ref{FCNC-box}.
%
For these box diagrams we obtain results 
\bea
({\rm \1})
\!\!&=&\!\!
\frac{-i g^4_{\rm EW}}{(4 \pi)^2 \cdot 4M^2_W} 
\sum_{i,j} \lambda_i \lambda_j {\cal F}^{(1)} (m_i,m_j,M_W,M_V,M_A) \times {\cal O}_{\Delta F=2}\,,
\label{box-gvm}
\\
({\rm \2})
\!\!&=&\!\!
\frac{-i g^4_{\rm EW}  X^2}{(4 \pi)^2 \cdot 4M^2_W}
\sum_{i,j} \lambda_i \lambda_j {\cal F}^{(2)} (m_i,m_j,M_W,\{ M \}) \times {\cal O}_{\Delta F=2}\,,
\\
\label{box-gvms}
({\rm \3})
\!\!&=&\!\!
\frac{-i g^4_{\rm EW} X^4}{(4 \pi)^2 \cdot 4M^2_W}  
\sum_{i,j} \lambda_i \lambda_j {\cal F}^{(3)} (m_i,m_j,M_W,\{  M \}) \times {\cal O}_{\Delta F=2}\,,
\eea
where 
$\{ M \}$ stands for the collection $\{ M_V,M_A,M_\pi \}$, the operator ${\cal O}_{\Delta F=2}$ is 
\bea
{\cal O}_{\Delta F=2} =
\begin{cases}
(\bar{s}_L \gamma^\mu d_L)(\bar{s}_L \gamma_\mu d_L) & \text{for $F=S$}\,, \\[2ex]
(\bar{b}_L \gamma^\mu q_L)(\bar{b}_L \gamma_\mu q_L) & \text{for $F=B_q\,,\,(q = d,s)$}\, .
\end{cases}
\eea
and $\lambda_i$ is given in terms of the CKM matrix elements by
\bea
\lambda_i =
\begin{cases}
V_{id} V^*_{is} & \text{for $F=S$}\,, \\[2ex]
V_{iq} V^*_{ib} & \text{for $F=B_q\,,\,(q = d,s)$}\,.
\end{cases}
\label{def-lambda_i}
\eea
The functions ${\cal F}^{(1,2,3)}$ are given explicitly in Appendix \ref{flavorfunctions}.

We have used the physical $SU(2)_L$ gauge coupling $g_{\rm EW}$, which is related to the bare gauge coupling $g$ as
\bea
\frac{1}{g^2_{\rm EW}} 
=
\frac{1}{g^2} \left[ 1+ \frac{1+(1-\chi)^2}{2} \epsilon^2 \right]
=
\frac{1}{g^2} \left[ 1+ \frac{1+(1-\chi)^2}{2} \bar{\epsilon}^2 \right]
\,,
\label{g-gew-relation}
\eea
where $\chi$ is a numerical factor introduced as in \cite{Appelquist:1999dq}, and whose value will be determined in Sec. {\ref{numresults}. 
This result is easily obtained by equating the well known Fermi coupling $G_F\sim g_{\rm EW}^2/M_W^2$ with the similar quantity $g^2/\tilde{M}_W^2$ which would be identified as the Fermi coupling in case this theory was applied as the Standard Model; see appendix \ref{flavorfunctions} for the expressions relating the parameters of the gauge and mass eigenbases. In the first equality we have used the definition $\epsilon=g/\tilde{g}$, where $\tilde{g}$ is the self coupling of the vector mesons, and expanded up to the order ${\cal O}({\epsilon}^4)$. The second equality is obtained with the definition $\bar{\epsilon}= g_{\rm EW}/\tilde{g}$, and noting that $\bar{\epsilon}^2=\epsilon^2$ up to contributions of order ${\cal O}(\bar{\epsilon}^4)$. For the results we present in this paper the quantity $\bar{\epsilon}$ is more convenient than $\epsilon$ since $g_{\rm EW}$ is the relevant coupling for the physical states. 

In Appendix \ref{flavorfunctions} we have collected the expressions relating the gauge and mass eigenbases up to ${\cal O}(\epsilon^5)$, but in the applications here, we will generally neglect ${\cal O}(\bar{\epsilon}^3)$ and higher terms in all expressions. 

In total, then, the $\Delta F = 2$ FCNC process is described by an effective Lagrangian given by
\bea
{\cal L}(\Delta F = 2) 
=
-\frac{G^2_F M^2_W}{4\pi^2} \cdot {\cal F}_0(m_i,m_j,M_W,M_V,M_A,M_\pi) \cdot {\cal O}_{\Delta F = 2}\,,
\eea 
where 
\bea
{\cal F}_0(m_i,m_j,M_W,\{ M \})
&=&
\sum_{i,j} \lambda_i \lambda_j
\left[ {\cal F}^{(1)} +X^2 {\cal F}^{(2)} +X^4 {\cal F}^{(3)} \right] \nonumber\\
&=&
{\cal F}^{\rm SM}_0(m_i,m_j,M_W) 
+
\Delta {\cal F}_0(m_i,m_j,M_W,\{ M \})\,.
\eea
Here $X$ is defined in (\ref{Xvariable}), and we have decomposed the result into a SM contribution, ${\cal F}^{\rm SM}_0 \equiv \sum \lambda_i \lambda_j {\cal F}_1(m_i,m_j,M_W)$, and a contribution arising solely from the new physics, 
\bea
\Delta {\cal F}_0=
\Delta {\cal F}^s_0(m_i,m_j,M_W,M_\pi)
+
\bar{\epsilon}^2 \Delta {\cal F}^{\rm vm}_0(m_i,m_j,M_W,M_V,M_A,M_\pi)
\,.
\label{f0def}
\eea
Furthermore, here $\Delta {\cal F}^s_0$ is the contribution solely from $\pi^\pm$ exchange 
and $\bar{\epsilon}^2 \Delta {\cal F}^{\rm vm}_0$ is the contribution from diagrams including vector mesons together with the charged pions. 

When we estimate the flavor observables in our setting, we take into account the unitarity of the CKM matrix by imposing $m_u \to 0$ \cite{Inami:1980fz}. For example in the case of $F = S$,
\bea
{\cal F}_0
=
\eta_1 \cdot \lambda^2_c \bar{{\cal F}}_0(m_c,M_W,\{ M \}) 
+
\eta_2 \cdot \lambda^2_t \bar{{\cal F}}_0(m_t,M_W,\{ M \})
+
\eta_3 \cdot 2 \lambda_c \lambda_t \bar{{\cal F}}_0(m_c,m_t,M_W,\{ M \})
\,,
\eea
where $\eta_{1,2,3}$ encode the QCD corrections for each $\bar{{\cal F}}_0$, and which are given by $\eta_1 = 1.44\,,\eta_2=0.57\,,\eta_3=0.47$ \cite{Nierste:2009wg}.
We note that, in the case of $F=B$ we need only $\eta_2 = \eta_B = 0.55$ \cite{Nierste:2009wg} because the top quark dominates the contributions to the box diagrams for $\Delta B=2$ processes. 
The function $\bar{{\cal F}}_0$ is given by
\bea
\bar{{\cal F}}_0(m_i,M_W,\{ M \})  = \lim_{m_u \to 0} \bar{{\cal F}}_0(m_u,m_i,M_W,\{ M\})\,,
\eea
and we use the charm quark mass $m_c = 1.3 \GeV$ \cite{UT-CKM}, and the top quark mass $m_t = 163.4 \GeV$ \cite{UT-CKM}.

In accordance with \cite{Bona:2007vi}, we define $C_{\epsilon_K,\Delta M_K, B_q}$ as
\bea
C_{\epsilon_K} 
&\equiv&
\frac{{\rm Im} \left[ \bra{K^0} {\cal H}^{\rm full}_{\rm eff} \ket{\bar{K^0}}\right]}{{\rm Im} \left[ \bra{K^0} {\cal H}^{\rm SM}_{\rm eff} \ket{\bar{K^0}} \right]} 
=
1 + \frac{{\rm Im}\left[ \Delta {\cal F}_0(F=S) \right]}{{\rm Im} \left[ {\cal F}^{\rm SM}_0 (F=S)\right]}\,,\label{c1def}
\\[1ex]
C_{\Delta M_K} 
&\equiv&
\frac{{\rm Re} \left[ \bra{K^0} {\cal H}^{\rm full}_{\rm eff} \ket{\bar{K^0}} \right]}{{\rm Re} \left[ \bra{K^0} {\cal H}^{\rm SM}_{\rm eff} \ket{\bar{K^0}}\right]} 
=
1 + \frac{{\rm Re}\left[ \Delta {\cal F}_0 (F=S)\right]}{{\rm Re}\left[ {\cal F}^{\rm SM}_0 (F=S)\right]}\,,
\\[1ex]
C_{B_q} 
&\equiv&
\left| 
\frac{\bra{B^0_q} {\cal H}^{\rm full}_{\rm eff} \ket{\bar{B^0_q}}}{\bra{B^0_q} {\cal H}^{\rm SM}_{\rm eff} \ket{\bar{B^0_q}}}
\right|
=
\left| 
1 + \frac{\Delta {\cal F}_0(F=B)}{{\cal F}^{\rm SM}_0(F=B)}
\right| \,,\label{c3def}
\eea
where ${\cal F}^{\rm SM}_0(F=S,B)$ are the SM contributions, ${\cal F}^{\rm SM}_0$ defined earlier, with $\lambda_i $ in Eq.(\ref{def-lambda_i}) chosen in correspondence with the process indicated in parentheses by $F=S$ or $F=B$. 

The UT-fit collaboration provides constraints on each $C$ parameter defined above; these are collected in Table \ref{ut-constraints-capri}.
\begin{table}
\begin{center}
\begin{tabular}{|c||c|c|}
\hline
\rule[-9pt]{0pt}{24pt}
parameter &  $68\%$ probability & $95\%$ probability
\\
\hline
\rule[-9pt]{0pt}{24pt}
$C_{\epsilon_K}$ & $1.09 \pm 0.13$ & [0.86, 1.39]
\\ \hline
\rule[-9pt]{0pt}{24pt}
$C_{\Delta M_K}$ & $1.17 \pm 0.42$ & [0.61, 2.43]
\\ \hline
\rule[-9pt]{0pt}{24pt}
$C_{B_d}$ & $0.97 \pm 0.14$ & [0.72, 1.30]
\\ \hline
\rule[-9pt]{0pt}{24pt}
$C_{B_s}$ & $0.96 \pm 0.10$ & [0.79, 1.18]
\\ \hline
\end{tabular}
\caption{
Constraints for $C_{\epsilon_K},C_{\Delta M_K}$ and $C_{B_q}$ \cite{UT-Capri2010}.
\label{ut-constraints-capri}
}
\end{center}
\end{table}
Finally, the CKM matrix is given by \cite{UT-CKM}
\bea
V_{\rm CKM}= 
\bpm
0.97427 
& 0.22535 
& 0.00377 
e^{i(-70.0^\circ) 
} 
\\
-0.22525 
e^{i (-0.0358^\circ) 
}
 & 0.97345
 & 0.04082 
\\
0.00869 
e^{i(-23.3^\circ)
} & -0.04007 
e^{i(-1.138^\circ) 
} & 0.99916 
\epm\,.
\eea

\section{Numerical results for Bosonic NMWT}
\label{numresults}
Let us start with the oblique corrections. As already mentioned, in our case, the $S$ parameter is given by
\bea
S = S_{V,A} + S_S\,,
\eea
where $S_{V,A}$ indicates the contribution from the vector mesons $V,A$ and 
$S_S$ indicates the contribution from the scalar mesons $\pi$ and the Higgs bosons. On the other hand, $S_{V,A}$ is related to the 0th Weinberg sum rule, because this sum rule is derived from vector-axial current correlation functions with vector meson saturation, and this does not have any contribution from physical scalar mesons. The 0th Weinberg sum rule is represented as \cite{Foadi:2007ue}
\bea
S_{V,A} 
= 4\pi \left[ \frac{F^2_V}{\tilde{M}^2_V} - \frac{F^2_A}{\tilde{M}^2_A}\right]
= \frac{8 \pi}{\tilde{g}^2} \left[ 1 - (1 - \chi)^2\right] +{\cal O}(\bar{\epsilon}^2) 
\,.
\label{0th-WSR}
\eea

The parameter $\chi$ is related to the parameters of the underlying theory, but is in principle free to take any value and hence $S_{V,A}$ can be vanishing or even negative. For our analysis, we will take $S_{V,A} = 0$, i.e. $\chi=1$. The calculation of $S_S$ was detailed in Sec. \ref{obliques}, and here we turn directly to the numerical results.

We demand that the values of $S$ and $T$ are within the 90\% confidence limit as
obtained by the PDG \cite{Amsler:2008zzb}. The resulting constraints for
the pion mass as well as the masses of the scalars are shown in figure
\ref{STconstraint}. As already noted in \cite{Antola:2009wq}, the
electroweak precision constraints favor a scenario of one light and one
heavy scalar particle, whereas the pion mass is not very strictly limited by the electroweak precision data.
\begin{figure}[h]
\begin{center}
\includegraphics[scale=0.4]{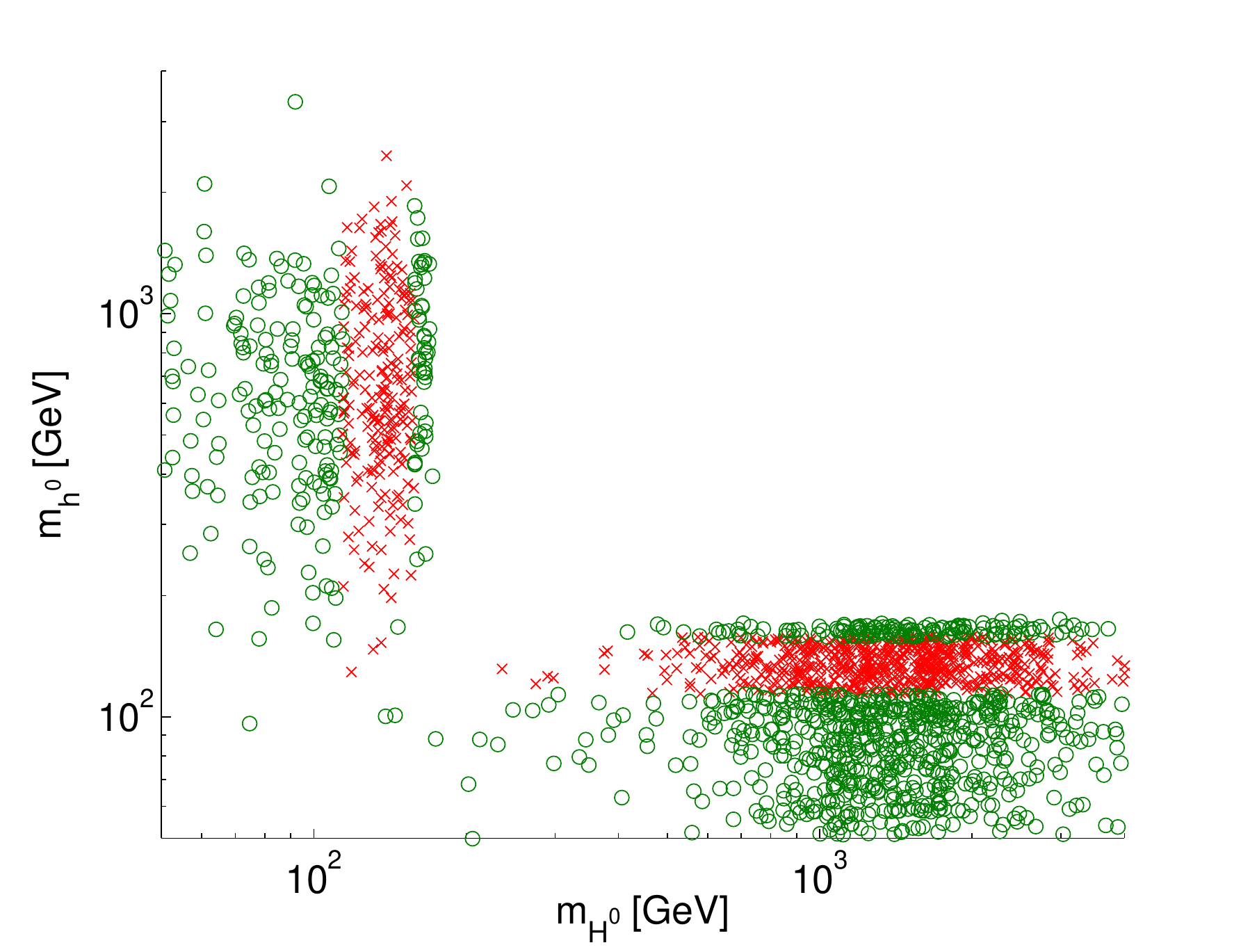}
\includegraphics[scale=0.4]{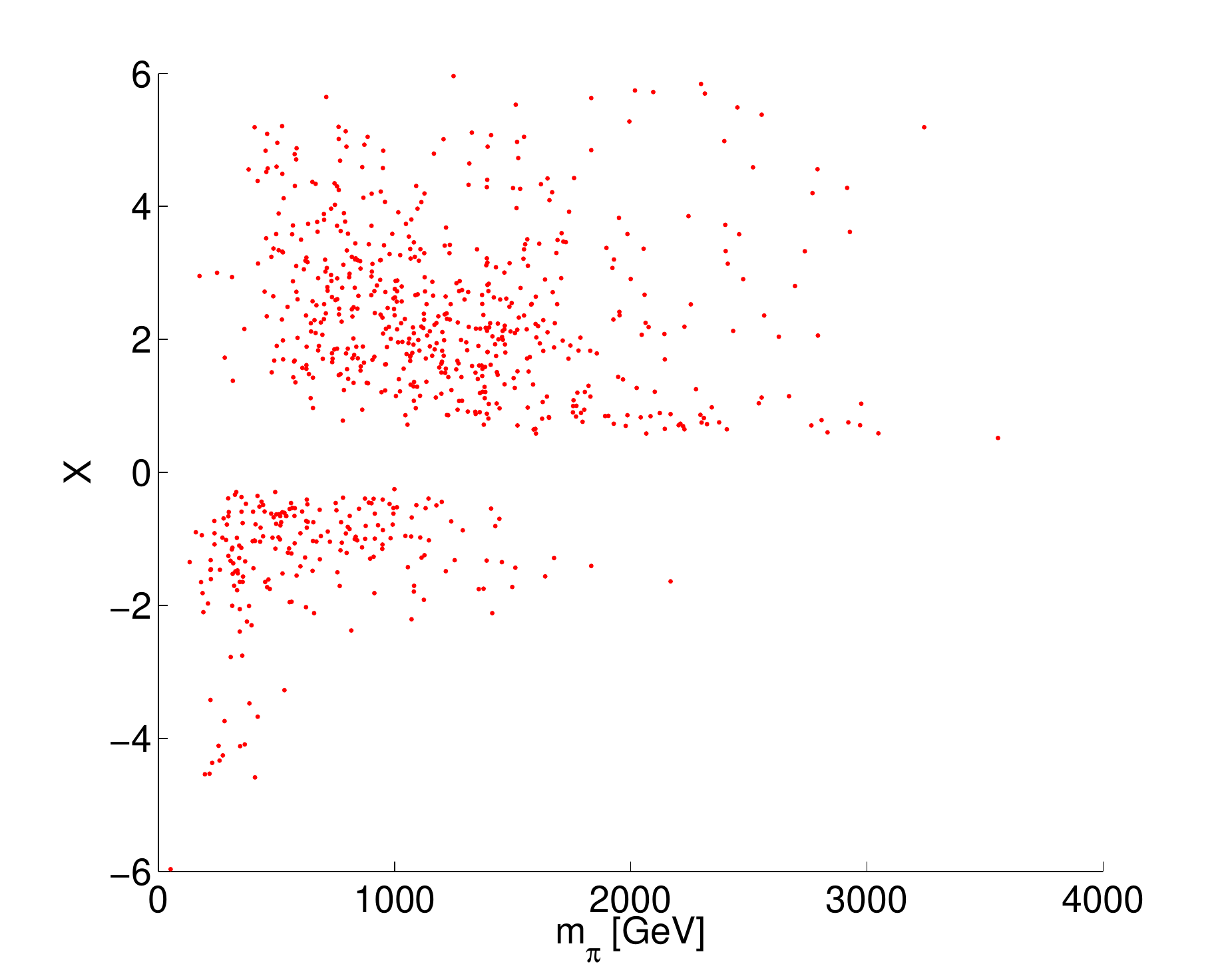}
\end{center}
\caption{Left panel: The masses of the scalars $h^0$ and $H^0$ after the
 constraints from $S$ and $T$ have been applied. The points marked by green circles are
 ruled out by direct detection limits from LEP and Tevatron, assuming
 SM-Higgs like branching ratios. The red crosses are allowed. Right
 panel: the mass of the pion and the variable $X$ as defined in equation
 (\ref{Xvariable}), after the electroweak precision constraints and the
 direct detection limits have been applied.
}
\label{STconstraint}
\end{figure}

Then we turn to the analysis of the flavor constraints. As detailed in Sec. \ref{flavor}, we have six parameters at our disposal. These are $X, M_V, M_A, M_\pi, \tilde{g}$ and $\chi$. To reduce the number of free parameters further we apply Weinberg sum rules. In addition to the 0th sum rule given above, the 1st Weinberg sum rule is given by \cite{Foadi:2007ue}
\bea
v^2_{\rm EW} = F^2_V - F^2_A 
\label{1WSR}
\,,
\eea
where $v_{\rm EW}$ satisfies $v^2_{\rm EW} = 4 \tilde{M}^2_W/\tilde{g}^2_{\rm EW} = 4 M^2_W/g^2_{\rm EW}$.
So the 1st Weinberg sum rule becomes
\bea
M^2_W = \dfrac{\epsilon^2}{2} \left[ M^2_V - (1-\chi)^2 M^2_A \right]\,,
\label{1WSR-mass}
\eea
where we neglect ${\cal O} (\bar{\epsilon}^4)$ contributions. 

Thus,  with the two Weinberg sum rules we reduce the parameters as follows:  With eqs.(\ref{0th-WSR}) and given $S_{V,A}=0$, we have $\chi=1$.
Then via (\ref{1WSR-mass}), we express $M_V$ in terms of the other parameters and this leaves $\tilde{g}, M_A, M_\pi$ and $X$ as free parameters. 
For our analysis, we choose certain benchmark values for $\tilde{g}$ and $M_A$ and then scan the $(m_\pi,X)$-plane. For each set of parameters we evaluate $\Delta{\cal F}$ as defined in (\ref{f0def}), and then use the constraints on (\ref{c1def})-(\ref{c3def}) to see if the corresponding point in the parameter space is viable.

Results of this procedure for $\chi=1$, $\tilde{g}=1,5$ and $M_A=400$ GeV or 600 GeV are shown in Fig.\ref{fig-UTfitconstraints}.
We checked that we cannot find any significant difference from the present results if the value of $S_{V,A}$ is varied from zero to $1/\pi$, where $S =1/\pi$ corresponds to the perturbative naive value of $S$ in the NMWT model. 
To comment about a difference in the flavor constraints arising from varying
$S=1/\pi$ to $S=0$ we note the following: The lower bound constraint in the figures coming from $C_{\epsilon_K}$ is slightly larger with $S=0$ than with $S=1/\pi$. On the other hand, the
upper constraint coming from $C_{B_s}$ is slightly smaller with $S=0$
than with $S=1/\pi$. These differences are only on the level of ${\cal O}(10^{-2} \%)$ for a fixed value of $m_\pi$, independently of the values of $\tilde{g},M_A$. 

However, the constraints on the $S$ and $T$ from precision data are more
stringent: allowing for $S_{V,A}\simeq 0.1$ in addition to the scalar
contribution would require the presence of a light scalar with mass
below 100 GeV which would be difficult to reconcile with
phenomenologically, even if its couplings were different from a Standard
Model Higgs. In general, the larger the value for $S_{V,A}$ is assumed,
the smaller the mass of the lighter one of the scalar boson states needs
to be. For example, if we take $S_{V,A}=0.05$, the parameter space
points where the lightest scalar mass is of the order of 170 GeV are
ruled out and we are left with points where the mass is of the order of
130 GeV or less. Thus, depending on the value of $S_{V,A}$, the parameter
space may be highly restricted by the combination of electroweak
precision observables (favoring the existence of a light scalar) and SM
Higgs searches. A similar tension on the Higgs mass due to precision
measurements and direct observation constraints is also present in the
SM itself, as well as in typical supersymmetric extensions of the SM.
\begin{figure}[h]
\begin{center}
\begin{tabular}[b]{cc}
{
\begin{minipage}[t]{0.5\textwidth}
\includegraphics[scale=0.35]{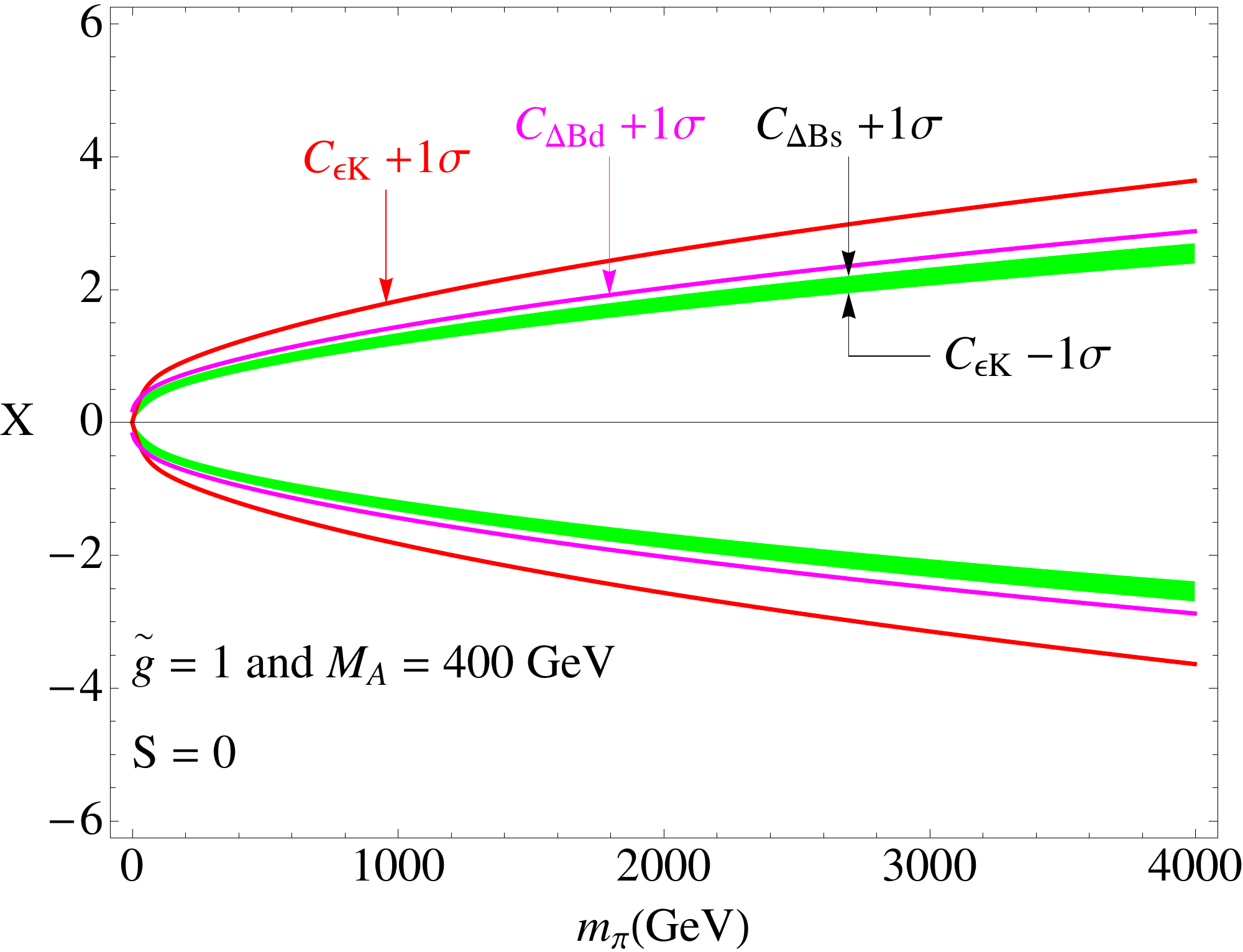}
\end{minipage}
}{
\begin{minipage}[t]{0.5\textwidth}
 \includegraphics[scale=0.35]{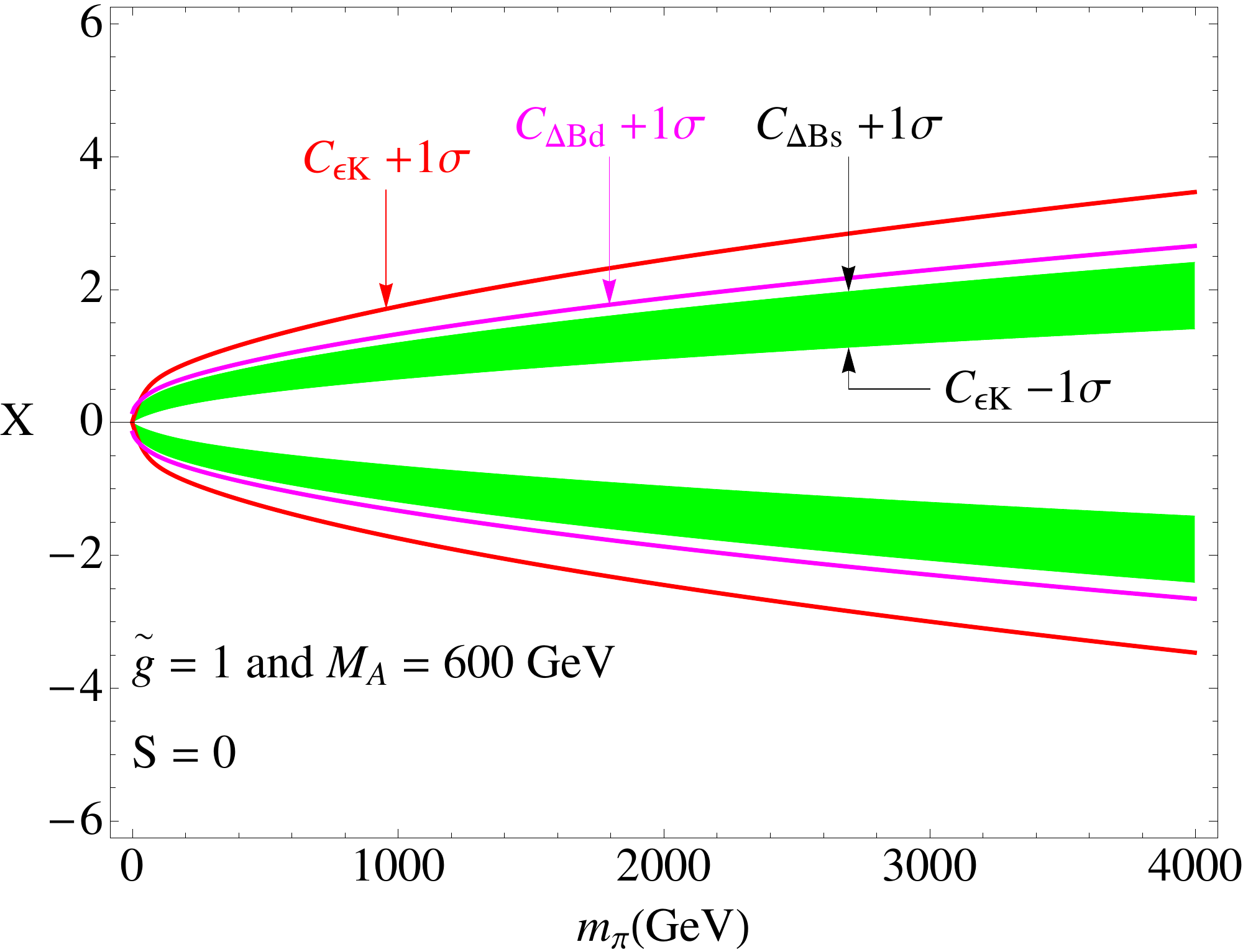}
\end{minipage}
}
 \\[1ex]
{
\begin{minipage}[t]{0.5\textwidth}
  \includegraphics[scale=0.35]{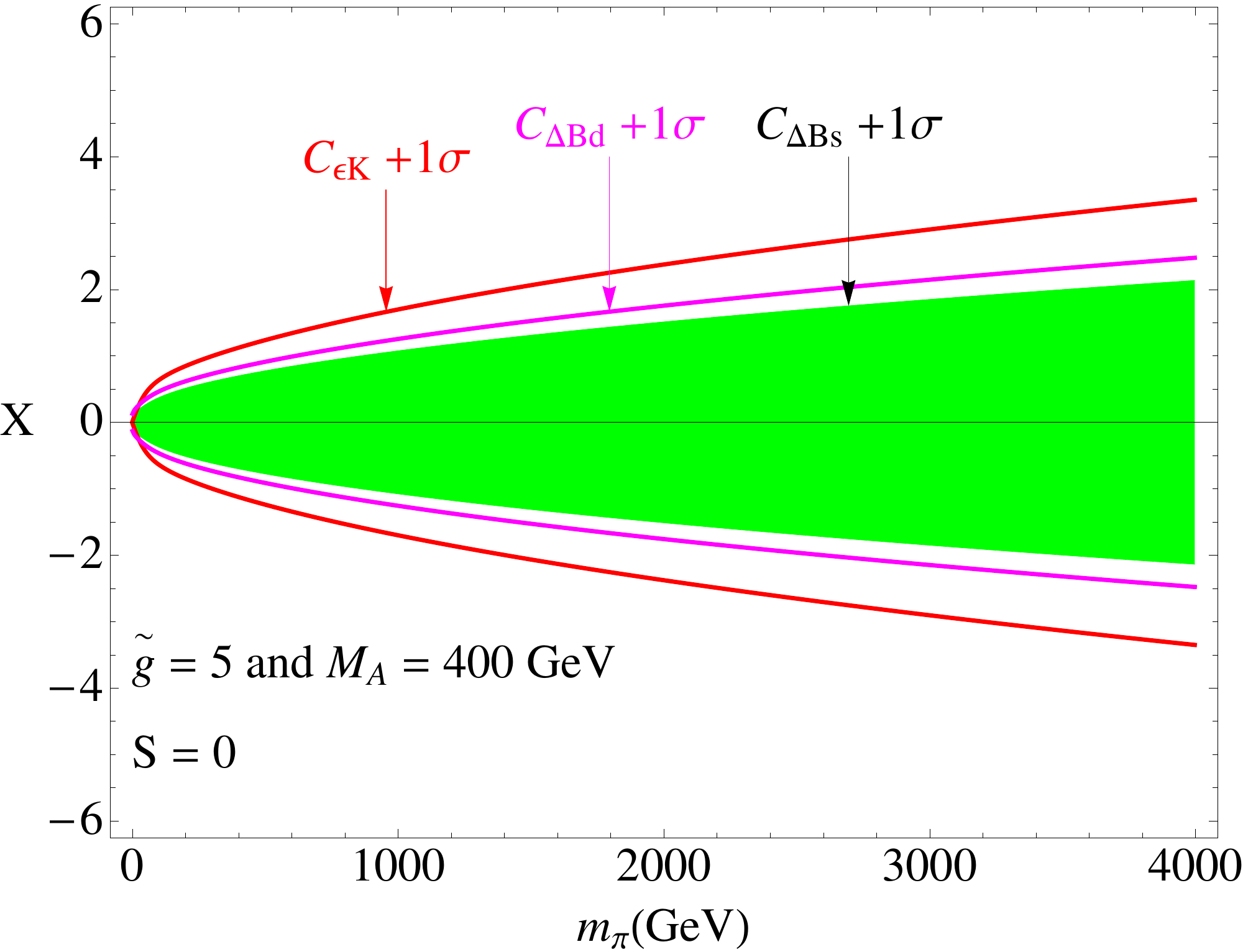}
\end{minipage}
}{
\begin{minipage}[t]{0.5\textwidth}
   \includegraphics[scale=0.35]{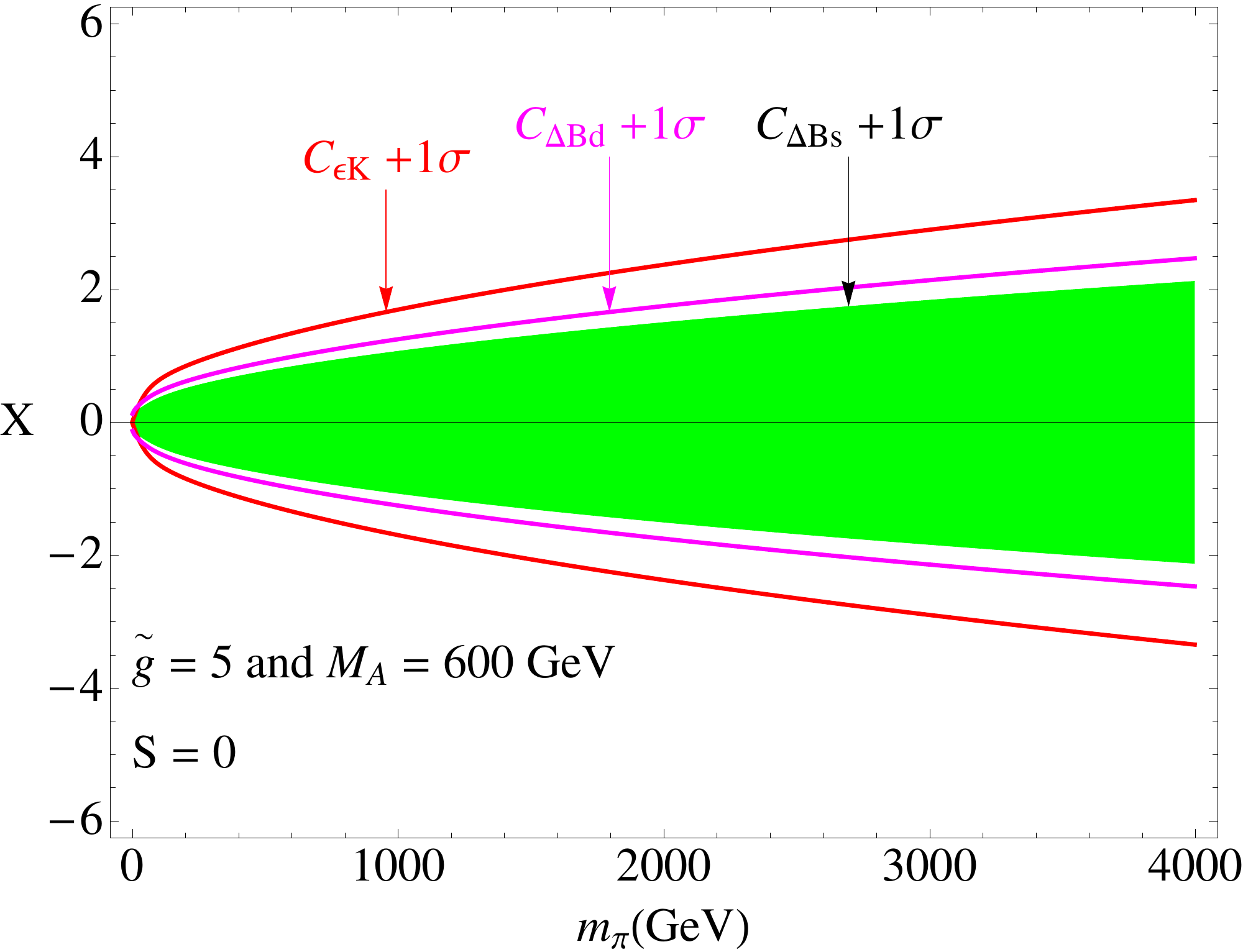}
\end{minipage}
}
\end{tabular}
\end{center}
\caption{
The UT-fit constraints coming from $C_{\epsilon_K,\Delta M_K, B_q}$ ($68\%$ probability) on $(M_\pi,X)$-plane for $(\tilde{g},M_A) = (1,400 \GeV)$ [top-left], $(1,600 \GeV)$ [top-right], $(5,400\GeV)$ [bottom-left] and $(5,600 \GeV)$ [bottom-right] with $S _{V,A}= 0$. The shaded region is allowed region. 
We show $C_{\epsilon_K,\Delta M_K, B_q}$ ($68\%$ probability) constraints in all panels.
In the bottom panels, the lower constraints coming from $C_{\epsilon_K}$ disappears. 
\label{fig-UTfitconstraints}
}
\end{figure}

\begin{figure}[h]
\begin{center}
\begin{tabular}[b]{cc}
{
\begin{minipage}[t]{0.5\textwidth}
\includegraphics[scale=0.35]{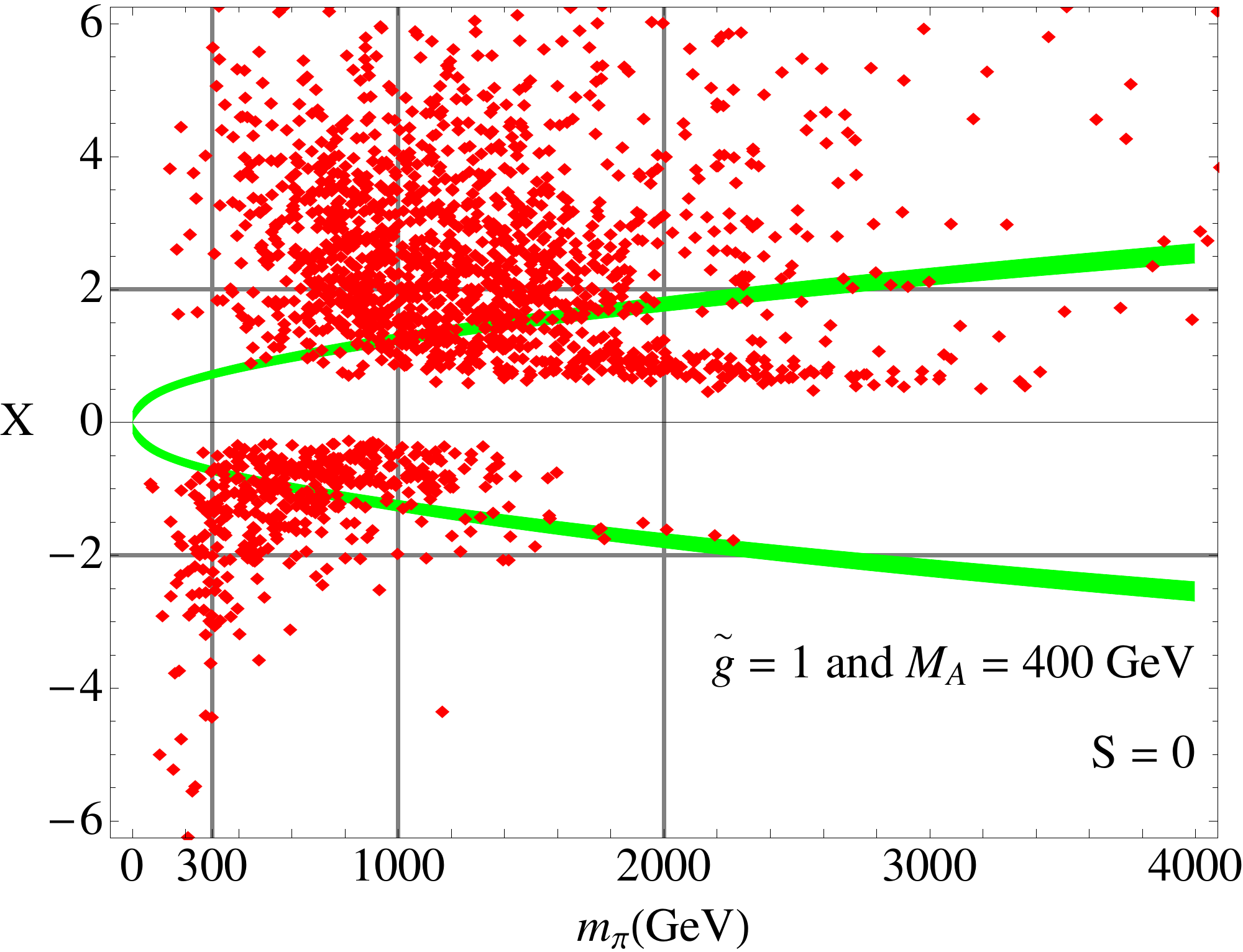}
\end{minipage}
}{
\begin{minipage}[t]{0.5\textwidth}
 \includegraphics[scale=0.35]{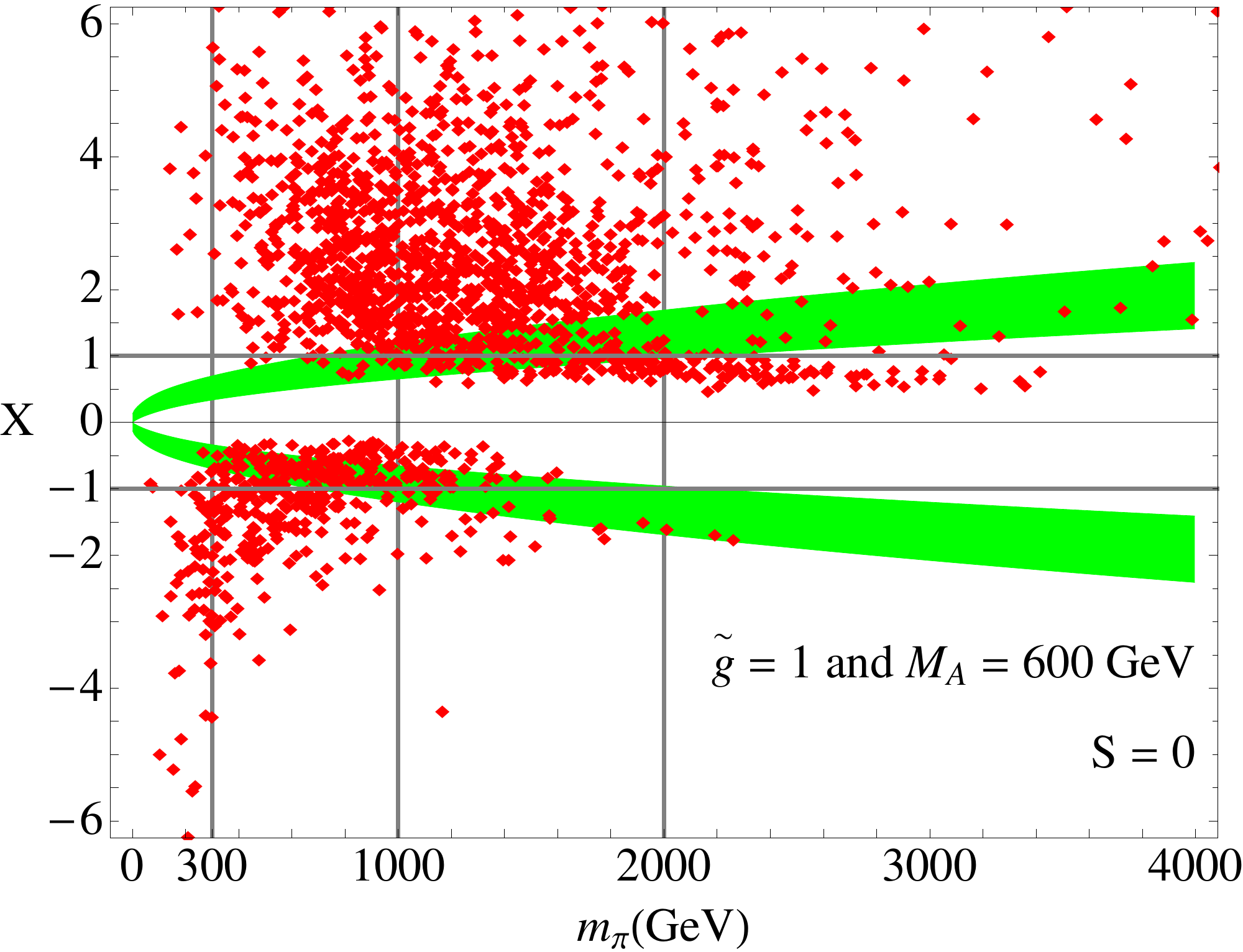}
\end{minipage}
}
 \\[1ex]
{
\begin{minipage}[t]{0.5\textwidth}
  \includegraphics[scale=0.35]{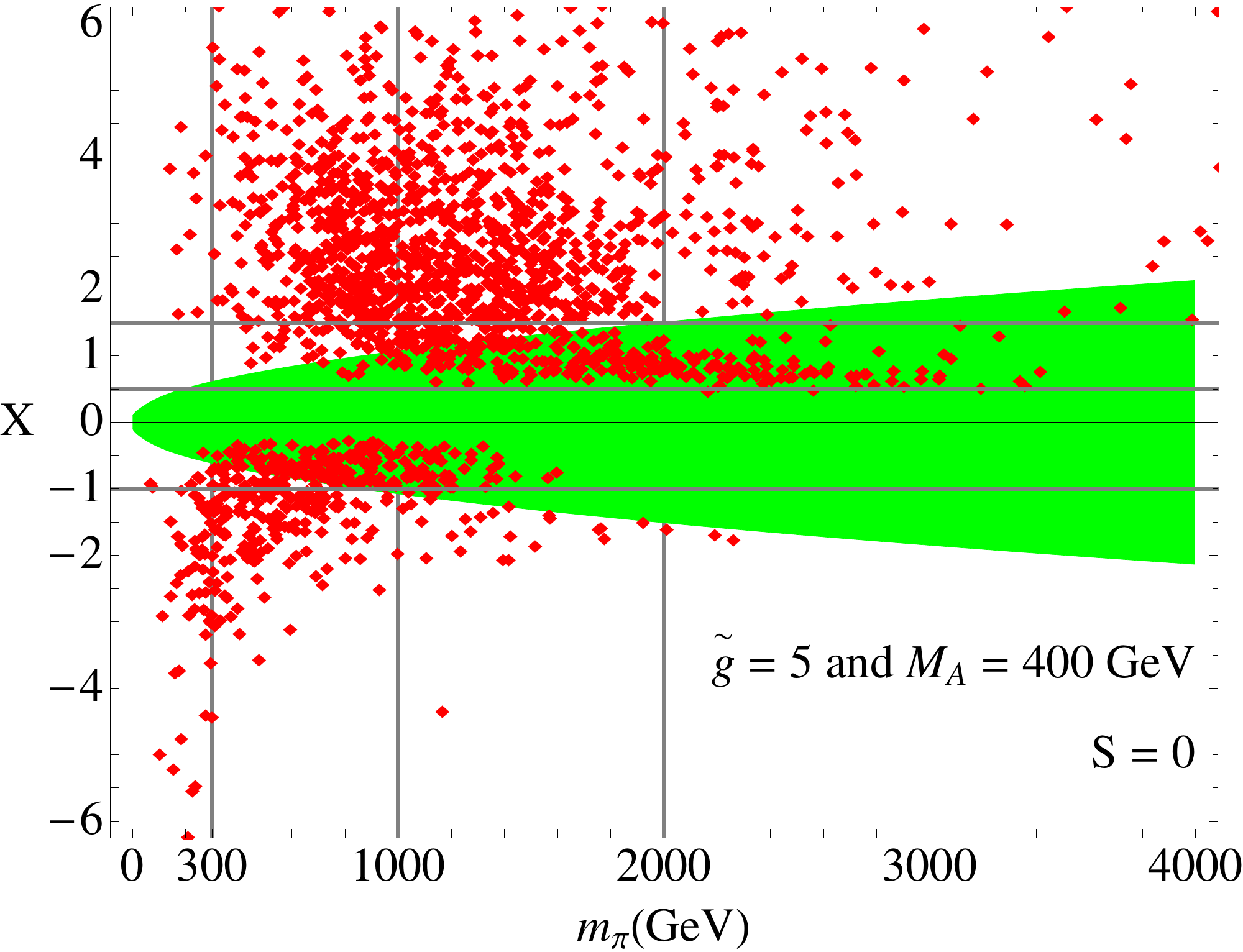}
\end{minipage}
}{
\begin{minipage}[t]{0.5\textwidth}
   \includegraphics[scale=0.35]{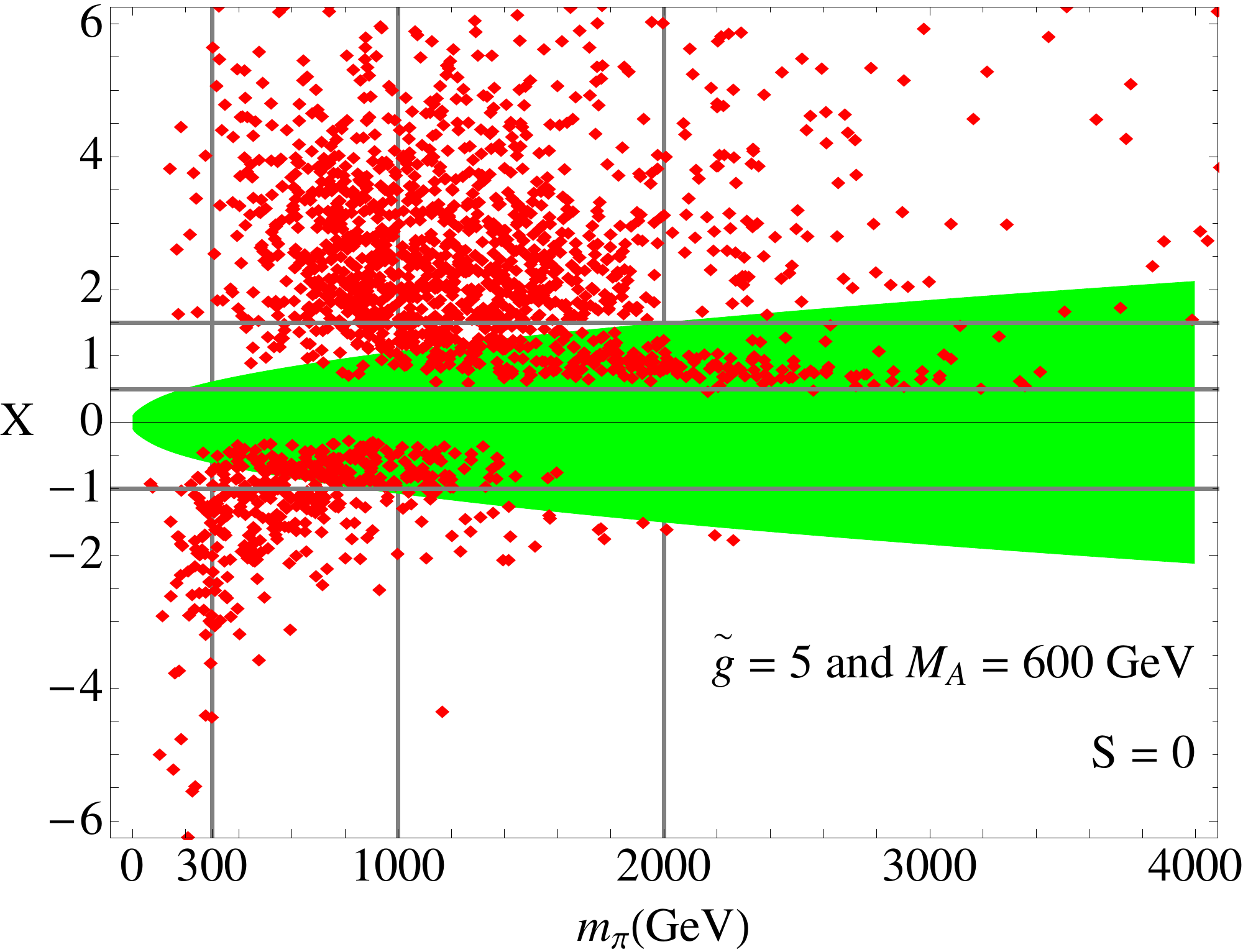}
\end{minipage}
}
\end{tabular}
\end{center}
\caption{
Constraints from flavor and electroweak oblique corrections on $(M_\pi,X)$-plane for $(\tilde{g},M_A) = (1,400 \GeV)$ [top-left], $(1,600 \GeV)$ [top-right], $(5,400\GeV)$ [bottom-left] and $(5,600 \GeV)$ [bottom-right] with $S _{V,A}= 0$.
The red diamonds correspond to the red crosses in the right panel in Fig.\ref{STconstraint}. 
The shaded regions correspond to the shaded region in Fig.\ref{fig-UTfitconstraints}.
\label{fig-UTconstraint-w-EWPT}
}
\end{figure}%
As one can see from Fig. \ref{fig-UTfitconstraints}, the most severe
upper bound comes from $C_{B_s}$ and the lower $68\%\cl$ constraint
comes from $C_{\epsilon_K}$ in table \ref{ut-constraints-capri} . A
lower constraint coming from $C_{\epsilon_K}$ disappears if $\tilde{g}
\geq 2$ or $M_A \geq 650 \GeV$. So, in this case it is enough to take
into consideration the  flavor constraint coming solely from the upper
$C_{B_s}$ constraint. In Fig.\ref{fig-UTconstraint-w-EWPT}, we show the
flavor constraints (shaded region) in Fig.\ref{fig-UTfitconstraints}
with the constraints (red diamonds) from the electroweak oblique
corrections for scalar sector in Fig.\ref{STconstraint}. Therefore our
analysis seems to indicate that electroweak oblique corrections and
flavor constraints taken together prefer 
\bea
\begin{cases}
 -1< X<0 \quad \text{and} \quad 300 \GeV < m_\pi < 1.6 \TeV
&
\text{for any $\tilde{g}$ and $M_A$}  \,,
\\[2ex]
\,1\,<\,X\,<\,2 \quad \text{and} \quad 800 \GeV < m_\pi < 2 \TeV
&
\text{for} \quad \tilde{g} \leq 2 \quad \text{and} \quad  M_A \leq 650 \GeV\,,
\\[2ex]
0.5<X<1.5 \quad \text{and} \quad 800 \GeV < m_\pi 
&
\text{for} \quad \tilde{g} \geq 2 \quad \text{or} \quad  M_A \geq 650 \GeV\,.
\end{cases}
\eea
In any of these cases, we see that it is possible that $m_\pi > M_A$ in the bosonic NMWT  model. This ordering in the spectrum would be different from the case of QCD-like theories.

\section{Conclusions and outlook}

In this paper we have revisited the bosonic next-to-minimal Walking Technicolor (NMWT) model introduced in \cite{Antola:2009wq}. We have extended the viability analysis of the model by considering the flavor constraints. The model, and the analysis carried out here, illustrate an important generic feature of flavor constraints in the Technicolor context: while there are the contributions due to the extension of the basic Technicolor theory, there are also intrinsic contributions arising from the Technicolor theory itself. These latter contributions originate from the vector mesons which mix with the electroweak gauge bosons. The former contributions were evaluated in this paper for bosonic Technicolor, but clearly similar analysis should be carried out for models where fermion mass generation is due to extended Technicolor (ETC) interactions.

As a result of our analysis we conclude that bosonic NMWT remains as a viable low energy model able to provide for masses of the SM matter fields. Of course the hierarchical mass patterns remain unexplained in the sense that they are merely parametrized in terms of Yukawa couplings similarly as e.g. in (MS)SM. Our results show that imposing the flavor constraints together with the constraints from oblique corrections provides bounds on the mixing patterns of the scalar states and masses of the physical pions.

\appendix
\section{Weak eigenstates}
\label{masseigenvalues}
In the NMWT model, mass term of the charged vector mesons is given by \cite{Foadi:2007ue}
\bea
{\cal L}_{CS} =
(\tilde{W}^-_\mu \, \tilde{V}^-_\mu \, \tilde{A}^-_\mu) \cdot \tilde{{\cal M}}^2_{CS} \cdot (\tilde{W}^{+\mu} \, \tilde{V}^{+\mu} \, \tilde{A}^{+\mu})^T\,.
\eea
Here, the tilde notation for the charged vector states indicate the fields in terms of the weak basis and
\bea
\tilde{{\cal M}}^2_{CS}
=
\bpm
\tilde{M}^2_W & -\dfrac{g}{\sqrt{2} \tilde{g}}\cdot \tilde{M}^2_V 
& -\dfrac{g}{\sqrt{2} \tilde{g}} \cdot \tilde{M}^2_A \cdot (1 -\chi) \\[3ex]
-\dfrac{g}{\sqrt{2} \tilde{g}} \cdot \tilde{M}^2_V & \tilde{M}^2_V & 0 \\[3ex]
-\dfrac{g}{\sqrt{2} \tilde{g}} \cdot \tilde{M}^2_A \cdot (1 -\chi) & 0 & \tilde{M}^2_A 
\epm\,,
\eea
where notations are according to \cite{Foadi:2007ue} except for $\tilde{M}_{W,V,A}$ which are given by 
\bea
\tilde{M}^2_W =\dfrac{g^2}{\tilde{g}^2}\left[  \left(\frac{1}{2} + \omega\right) \tilde{M}^2_V - \frac{(1-\chi)^2}{2}\tilde{M}^2_A \right]
\quad,\quad 
\tilde{M}^2_V = \frac{\tilde{g}^2}{2} F^2_V
\quad , \quad 
\tilde{M}^2_A = \frac{\tilde{g}^2}{2(1-\chi)^2} F^2_A \,.
\eea
The parameters  $\chi$ and $\omega$ are related to the couplings in the underlying Lagrangian and like for $F_{V(A)}$, the (axial)vector-decay constant, their definitions are same as in \cite{Foadi:2007ue}.
The mass matrix $\tilde{\cal M}^2_{CS}$ is diagonalized by the orthogonal transformation 
and as a result we obtain the eigenvalues
$M^2_{W,V,A}$ which are given by ($\epsilon \equiv g/\tilde{g}$)
\bea
 M^2_W & = &
\tilde{M}^2_W \left[ 1 - \frac{1+(1-\chi)^2}{2} \cdot \epsilon^2 \right] \,,
\label{eigenvalue-MW}\\[1ex]
 M^2_{V^\pm}  & = &
\tilde{M}^2_V 
\left[ 1 + \frac{\tilde{M}^2_W + \tilde{M}^2_V }{2\tilde{M}^2_V} \cdot \epsilon^2 - \frac{(1-\chi)^2 \tilde{M}^2_A}{4(\tilde{M}^2_A - \tilde{M}^2_V)} \cdot \epsilon^4 \right]
 \,,
\label{eigenvalue-MV}\\[1ex]
M^2_{A^\pm}  & = &
\tilde{M}^2_A 
\left[ 1 + (1-\chi)^2\frac{\tilde{M}^2_W + \tilde{M}^2_A }{2\tilde{M}^2_A} \cdot \epsilon^2 - \frac{(1-\chi)^2 \tilde{M}^2_V}{4(\tilde{M}^2_V - \tilde{M}^2_A)} \cdot \epsilon^4\right] \,,
\label{eigenvalue-MA}
\eea
where we neglect ${\cal O}(\epsilon^5)$-terms.
Moreover, we present the mass eigenvectors 
as
\bea
\vec{v}_W 
= \bpm x_W \\ y_W \\ z_W \epm 
\quad , \quad 
 \vec{v}_V 
= \bpm x_V \\ y_V \\ z_V \epm 
\quad , \quad
\vec{v}_A 
= \bpm x_A \\ y_A \\ z_A \epm 
\,, \label{eigen-v}
\eea
in which each $x,y,z$ is given by
\bea
x_W
&=& 
1 
-\frac{1}{2} \cdot \dfrac{1+(1-\chi)^2}{2} \epsilon^2 
+ \frac{3}{8}\cdot \dfrac{[1+(1-\chi)^2 ]^2}{4} \epsilon^4 \nonumber\\
&& \hspace*{25ex}
- \dfrac{1}{2} \left[ 
\dfrac{\tilde{M}^2_W}{\tilde{M}^2_V} + (1-\chi)^2 \dfrac{\tilde{M}^2_W}{\tilde{M}^2_A} \right] \epsilon^2
\,,\label{xW}\\[1ex]
y_W
&=&
\dfrac{\epsilon}{\sqrt{2}}
\left[
1+ \dfrac{\tilde{M}^2_W}{\tilde{M}^2_V} - \dfrac{1+(1-\chi)^2}{4} \epsilon^2
 \right]
\,,\label{yW}\\[2ex]
z_W
&=&
\dfrac{(1-\chi)\epsilon}{\sqrt{2}} 
\left[
1+  \dfrac{\tilde{M}^2_W}{\tilde{M}^2_A} - \dfrac{1+(1-\chi)^2}{4} \epsilon^2
\right] 
\,,\label{zW}
\eea
\bea
x_V
&=&
-\dfrac{\epsilon}{\sqrt{2}} 
\left[ 1 + \dfrac{\tilde{M}^2_W}{\tilde{M}^2_V} - \frac{1}{2} \left(\dfrac{1}{2} + \dfrac{(1-\chi)^2 \tilde{M}^2_A }{\tilde{M}^2_A - \tilde{M}^2_V}\right) \epsilon^2
\right] \,,\\[1ex]
y_V
&=&
1 - 
\dfrac{\epsilon^2}{4} 
\left[ 
1 + 2\dfrac{\tilde{M}^2_W}{\tilde{M}^2_V} - \dfrac{1}{2} \left( \dfrac{3+ 4(1-\chi)^2}{4} - \dfrac{(1-\chi)^2 \tilde{M}^4_V }{( \tilde{M}^2_A - \tilde{M}^2_V )^2}\right)\epsilon^2 
\right]
\,,\\[1ex]
z_V
&=&
\dfrac{\epsilon^2 \cdot (1-\chi)\tilde{M}^2_A}{2(\tilde{M}^2_V - \tilde{M}^2_A)}  
\left[ 
1 + \dfrac{\tilde{M}^2_W}{\tilde{M}^2_V} - \dfrac{1}{2} \left( \dfrac{1}{2} - \dfrac{\tilde{M}^2_V - (1-\chi)^2 \tilde{M}^2_A}{\tilde{M}^2_A - \tilde{M}^2_V}\right) \epsilon^2
\right]
\,,
\eea
\bea
x_A
&=&
-\dfrac{(1-\chi)\epsilon}{\sqrt{2}} 
\left[ 1 + \dfrac{\tilde{M}^2_W}{\tilde{M}^2_A} -\frac{1}{2}\left( \dfrac{(1-\chi)^2}{2} + \dfrac{\tilde{M}^2_V }{\tilde{M}^2_V - \tilde{M}^2_A}\right) \epsilon^2
\right]
\,,\\[1ex]
y_A
&=&
\dfrac{\epsilon^2 \cdot (1-\chi)\tilde{M}^2_V}{2\left( \tilde{M}^2_A - \tilde{M}^2_V \right)}
\left[ 1+ \dfrac{\tilde{M}^2_W}{\tilde{M}^2_A} - \frac{1}{2}
\left( \dfrac{(1-\chi)^2}{2} + \dfrac{\tilde{M}^2_V - (1-\chi)^2 \tilde{M}^2_A}{\tilde{M}^2_V - \tilde{M}^2_A}\right) \epsilon^2
\right]
\,,\\[1ex]
z_A
&=&
1 - \dfrac{(1-\chi)^2 \epsilon^2}{4}  
\left[ 1 + 2  \dfrac{\tilde{M}^2_W}{\tilde{M}^2_A} -\frac{1}{2}\left( \dfrac{4+ 3(1-\chi)^2}{4} -\dfrac{\tilde{M}^4_A}{(\tilde{M}^2_V - \tilde{M}^2_A)^2}\right) \epsilon^2 
\right]\,,
\eea
where we neglect ${\cal O}(\epsilon^5)$ again. 

In the text we keep terms only up to ${\cal O}(\epsilon^3)$ for each estimation.

%
%

\section{The box diagrams}
\label{flavorfunctions}
The functions arising from the evaluation of the box diagrams are defined 
as
\bea
{\cal F}^{(1)}(m_i,m_j,M_W,M_V,M_A) 
\!\!&=&\!\! 
\bar{{\cal F}}_1(m_i,m_j,M_W) \nonumber\\[1ex]
&&
- \bar{\epsilon^2}
\left[ 
\begin{aligned}
&
2 \left( \dfrac{M^2_W}{M^2_V} + (1-\chi)^2 \dfrac{M^2_W}{M^2_A}\right)
\bar{{\cal F}}_1(m_i,m_j,M_W) 
\\[1ex]
&+
\frac{(M^2_V + M^2_W)^2}{M^4_V}
\bar{{\cal F}}_1(m_i,m_j,M_W,M_V) 
\\[1ex]
&+
(1-\chi)^2 \frac{(M^2_A + M^2_W)^2}{M^4_A}
\bar{{\cal F}}_1(m_i,m_j,M_W,M_A)
\end{aligned}
\right] \,,\nonumber \\
\eea
\bea
{\cal F}^{(2)}(m_i,m_j,M_W,\{ M \}) 
\!\!&=&\!\! 
2 \bar{{\cal F}}_2(m_i,m_j,M_W,M_\pi) \nonumber\\[1ex]
&&
- \bar{\epsilon^2}
\left[ 
\begin{aligned}
&
2 \left( \dfrac{M^2_W}{M^2_V} + (1-\chi)^2 \dfrac{M^2_W}{M^2_A}\right)
\bar{{\cal F}}_2(m_i,m_j,M_W,M_\pi) 
\\[1ex]
&-
\frac{(M^2_V + M^2_W)^2}{M^4_V}
\bar{{\cal F}}_2(m_i,m_j,M_V,M_\pi) 
\\[1ex]
&-
(1-\chi)^2 \frac{(M^2_A + M^2_W)^2}{M^4_A}
\bar{{\cal F}}_2(m_i,m_j,M_A,M_\pi)
\end{aligned}
\right] \,,\nonumber \\
\eea
\bea
{\cal F}^{(3)}(m_i,m_j,M_W,M_\pi) 
=
\frac{m^2_im^2_j}{M^2_W} A_2(m_i,m_j,M_\pi)\,,
\eea
and 
\bea
\frac{1}{M^2_W}\bar{{\cal F}}_1(m_1,m_2,M_1,M_2)
\!\!&=&\!\!
\left( 1 + \frac{m^2_1 m^2_2}{4M^2_1M^2_2} \right) A_2(m_1,m_2,M_1,M_2) 
\nonumber\\[1ex]
&&\hspace*{10ex} 
+ \frac{m^2_1m^2_2(M^2_1 + M^2_2)}{M^2_1M^2_2} A_0(m_1,m_2,M_1,M_2)\,, 
\\
\frac{1}{M^2_W}\bar{{\cal F}}_1(m_1,m_2,M)
&=&
\left( 1 + \frac{m^2_1 m^2_2}{4M^4} \right) A_2(m_1,m_2,M) 
+ \frac{2 m^2_1 m^2_2}{M^2}A_0(m_1,m_2,M)\,,\nonumber\\
\\
\bar{{\cal F}}_2(m_1,m_2,M_1,M_2)
\!\!&=&\!\!
m^2_1 m^2_2
\left[ \frac{1}{4M^2_W} A_2(m_1,m_2,M_1,M_2) 
+ A_0(m_1,m_2,M_1,M_2)
\right] \,. \nonumber
\\
\eea
$A_0$ and $A_2$ are given by 
\bea
A_0(m_1,m_2,M_1,M_2) 
\!\!&=&\!\!
\frac{-1}{m^2_1 - m^2_2}
\left[
\frac{m^2_1 \ln m^2_1}{(M^2_1-m^2_1)(M^2_2-m^2_1)}
-
\frac{m^2_2 \ln m^2_2}{(M^2_1-m^2_2)(M^2_2-m^2_2)}
\right] \nonumber\\[1ex]
&+&\!\!
\frac{-1}{M^2_1 - M^2_2}
\left[
\frac{M^2_1 \ln M^2_1}{(m^2_1-M^2_1)(m^2_2-M^2_1)}
-
\frac{M^2_2 \ln M^2_2}{(m^2_1-M^2_2)(m^2_2-M^2_2)}
\right] \,,\nonumber \\
\\
A_0(m_1,m_2,M) 
\!\!&=&\!\!
\frac{-1}{m^2_1 - m^2_2}
\left[
\frac{m^2_1 \ln \dfrac{m^2_1}{M^2}}{(M^2-m^2_1)^2}
-
\frac{m^2_2 \ln \dfrac{m^2_2}{M^2}}{(M^2-m^2_2)^2}
+
\frac{1}{M^2-m^2_1}
-
\frac{1}{M^2-m^2_2}
\right] \,,\nonumber\\
\\
A_0(m,M) 
\!\!&=&\!\!
\frac{-1}{(M^2 - m^2)^2}
\left[
\frac{(M^2+m^2) \ln \dfrac{m^2}{M^2}}{M^2-m^2}
+
2
\right] \,,
\eea
\bea
A_2(m_1,m_2,M_1,M_2) 
\!\!&=&\!\!
\frac{1}{m^2_1 - m^2_2}
\left[
\frac{m^4_1 \ln m^2_1}{(M^2_1-m^2_1)(M^2_2-m^2_1)}
-
\frac{m^4_2 \ln m^2_2}{(M^2_1-m^2_2)(M^2_2-m^2_2)}
\right] \nonumber\\[1ex]
&+&\!\!
\frac{1}{M^2_1 - M^2_2}
\left[
\frac{M^4_1 \ln M^2_1}{(m^2_1-M^2_1)(m^2_2-M^2_1)}
-
\frac{M^4_2 \ln M^2_2}{(m^2_1-M^2_2)(m^2_2-M^2_2)}
\right] \,,\nonumber \\
\\
A_2(m_1,m_2,M) 
\!\!&=&\!\!
\frac{1}{m^2_1 - m^2_2}
\left[
\frac{m^4_1 \ln \dfrac{m^2_1}{M^2}}{(M^2-m^2_1)^2}
-
\frac{m^4_2 \ln \dfrac{m^2_2}{M^2}}{(M^2-m^2_2)^2}
+
\frac{M^2}{M^2-m^2_1}
-
\frac{M^2}{M^2-m^2_2}
\right] \,,\nonumber\\
\\
A_2(m,M) 
\!\!&=&\!\!
\frac{1}{(M^2 - m^2)^2}
\left[
\frac{2 m^2 M^2 \ln \dfrac{m^2}{M^2}}{M^2-m^2}
+
m^2+M^2
\right] \,.
\eea


\end{document}